\documentclass[manuscript]{aastex61}
 \UseRawInputEncoding

\begin{document}

\title{The first flare observation with a new solar microwave spectrometer working in 35 - 40~GHz}

\author[0000-0002-4451-7293]{Fabao Yan}\altaffiliation{These authors contributed equally to this work}
\affil{Laboratory for Electromagnetic Detection, Institute of Space Sciences, Shandong University, Weihai, Shandong 264209, China;}

\author[0000-0002-6985-9863]{Zhao Wu}\altaffiliation{These authors contributed equally to this work}
\affil{Laboratory for Electromagnetic Detection, Institute of Space Sciences, Shandong University, Weihai, Shandong 264209, China;}

\author[0000-0002-8826-8006]{Ziqian Shang}
\affil{Laboratory for Electromagnetic Detection, Institute of Space Sciences, Shandong University, Weihai, Shandong 264209, China;}

\author{Bing Wang}
\affil{Laboratory for Electromagnetic Detection, Institute of Space Sciences, Shandong University, Weihai, Shandong 264209, China;}

\author{Lei Zhang}
\affil{Laboratory for Electromagnetic Detection, Institute of Space Sciences, Shandong University, Weihai, Shandong 264209, China;}

\author[0000-0001-6449-8838]{Yao Chen}\altaffiliation{Corresponding author: yaochen@sdu.edu.cn}
\affil{Laboratory for Electromagnetic Detection, Institute of Space Sciences, Shandong University, Weihai, Shandong 264209, China;}
\affil{Institute of Frontier and Interdisciplinary Science, Institute of Space Sciences,Shandong University; yaochen@sdu.edu.cn}

\begin{abstract}
The microwave spectrum contains valuable information about solar flares. Yet, the present spectral coverage is far from complete and broad data gaps exist above 20 GHz. Here we report the first flare (the X2.2 flare on 2022 April 20) observation of the newly-built Chashan Broadband Solar millimeter spectrometer (CBS) working from 35 to 40~GHz. We use the CBS data of the new Moon to calibrate, and the simultaneous NoRP data at 35~GHz to cross-calibrate. The impulsive stage has three local peaks with the middle one being the strongest and the maximum flux density reaches $\sim$9300 SFU at 35 - 40 GHz. The spectral index of the CBS data ($\alpha\rm{_C}$) for the major peak is mostly positive, indicating the gyrosynchrotron turnover frequency ($\nu_t$) goes beyond 35 - 40 GHz. The frequency $\nu_t$ is smaller yet still larger than 20 GHz for most time of the other two peaks according to the spectral fittings with NoRP-CBS data. The CBS index manifests the general rapid-hardening-then-softening trend for each peak and gradual hardening during the decay stage, agreeing with the fitted optically-thin spectral index ($\alpha_{tn}$) for $\nu_t < 35$ GHz. In addition, the obtained turnover frequency ($\nu_t$) during the whole impulsive stage correlates well with the corresponding intensity ($I_t$) according to a power-law dependence ($I_t \propto \nu_t^{4.8}$) with a correlation coefficient of 0.82. This agrees with earlier studies on flares with low turnover frequency ($\le 17$ GHz), yet being reported for the first time for events with a high turnover frequency ($\ge 20$ GHz).
\end{abstract}

\keywords{Sun: flares, Sun: radio radiation, Instrumentation: spectrographs, radiation mechanisms: non-thermal}
\section{Introduction}

Microwave emission of solar flares can be excited by energetic electrons through the gyrosynchrotron (GS) emission \citep[e.g., ][]{Ramaty1969,Dulk1985,Bastian1998,White2011,Nagnibeda2013,Nindos2020}. The typical microwave spectra peak below or around 10~GHz, at which the spectral slope changes from positive (corresponding to the optically-thick regime) to negative (the optically-thin regime). The spectral parameters, including the turnover frequency, the spectral index, and the flux density, can be used to infer the flaring process and conditions \citep[e.g., ][]{Petrosian1981,Dulk1982,Klein1987,Fleishman2010,Kuznetsov2011,Gary2013,Wu2016,Chen2017,Casini2017,Chen2020}.

According to a statistical study of strong microwave bursts using the Owens Valley Solar Array (OVSA) data, the turnover frequency is highly correlated with the flux density \citep{Melnikov2008,Lysenko2018}. Earlier authors limited their studies to events with smaller turnover frequency, either being $<$9.4~GHz \citep{Ning2007} or $<$17~GHz \citep{Asai2013}, to ensure two data points are available in the optically-thin regime. Yet, the turnover frequency can go beyond 20~GHz around the flare peak time. In addition, spectra with unusual shapes, such as flat or continuous rising even at frequencies above tens of GHz, have been reported \citep{White1992,Ramaty1994,Kaufmann2004,Silva2007,Zhou2010,Nagnibeda2013,Song2016}. \citet{Wu2019} found with GS simulations that the turnover frequency can reach up to tens of GHz, being sensitive to the abundance of energetic electrons of hundreds keV to a few MeV. These studies indicate the necessity of a full spectral coverage of the centimeter-millimeter wavelength to better understanding solar flares.

Present solar microwave spectrometers, including the Expanded OVSA \citep[1~-~18~GHz,][]{Gary2018}, the Siberian spectropolarimeter (2~-~24~GHz), and the Mingantu Ultrawide Spectral Radioheliogragh \citep[MUSER, 0.4~-~15 GHz,][]{Yan2009,Wang2013}, provide dynamic spectrum below $\sim$20 GHz. Above that, data exist only at a few discrete frequencies, e.g., the Nobeyama Radiopolarimeter \citep[NoRP,][]{Nakajima1985} measures the flux density at 35 and 80~GHz (the 80~GHz data have not been updated since 2015 according to the NoRP website\footnote{$ftp://solar-pub.nao.ac.jp/pub/nsro/norp/xdr/$\label{fn:1}}), the Mets\"{a}hovi Radio Observatory (MRO) images the Sun at 37~GHz. In addition, the Atacama Large Millimeter/submillimeter Array \citep[ALMA,][]{Wootten2009} works at 100 and 239~GHz \citep{Shimojo2017a,White2017}, and the Solar Submillimeter Telescope \citep[SST,][]{Kaufmann2001} at 212 and 405~GHz. Significant data gaps exist in the millimeter wavelength that corresponds to the optically-thin regime of most flares, thus this part for the flare physics has not been fully explored.

The newly-built Chashan Broadband Solar millimeter spectrometer (CBSmm, CBS for short) started its routine observation since 2020, working from 35 to 40~GHz \citep{Shang2022}. It is operated by the Institute of Space Sciences of Shandong University. On 2022 April 20, the microwave burst during an X2.2 flare was observed by both NoRP and CBS. This provides the first flare observation of CBS since its routine operation, with a good opportunity of cross-calibration with NoRP and further study of a major flare.

\section{Instruments and data calibration}
\label{section:2}
\subsection{Brief introduction to the CBS}
The CBS consists of three modules. The receiving module has a Cassegrainian antenna with a diameter of 80~cm. The analog front end (AFE) unit down-converts the receiving signal (35 - 40~GHz) to 687.5 - 1187.5 MHz into 10 channels. Each channel has a bandwidth of 500~MHz. The digital receiver generates the real-time dynamic spectra in two channels, with a dual-core analog-to-digital-converter (ADC) and a compatible processor of the field programmable gate array (FPGA). For details see \citet{Yan2020}, \citet{Yan2021}, and \citet{Shang2022}. To increase the signal-to-noise ratio (SNR) and reduce the data volume, we integrated the spectra over $\sim$134~ms that represents the time resolution of the data. The spectral resolution is $\sim153$~kHz.

\subsection{Data calibration}
The CBS data are calibrated with the CBS measurement of the new Moon on the basis of known brightness temperature ($T_{NM}(\nu)$) as follows
\begin{equation}
T_S(\nu)=\frac{R_{S}^F-R_{BG}^F}{R_{QS}^F-R_{BG}^F}\beta T_{NM}(\nu),\quad\beta =\frac{R_{QS}^{NM}-R_{BG}^{NM}}{R_M^{NM}-R_{BG}^{NM}},
\label{eq:calibration}
\end{equation}
where F (NM) denotes the flare (new Moon) observation, $R$ is the reading value of the receiver, $T_S(\nu)$ is the mean brightness temperature of the Sun, and BG, M, and QS denote the sky background, the Moon, and the quiet Sun, respectively. Given the weak polarization at 35~GHz (\textless3\%, see the blue line in Figure \ref{fig3}(e)), we assume equal $T_S(\nu)$ for the left- and right-handed circular polarizations. The total flux density $I(\nu)$ is then given by \citep{Dulk1982}
\begin{equation}
I(\nu)=\frac{2k_B\nu^2\int{T_S(\nu)d\Omega}}{c^2},
\label{eq:totalflux}
\end{equation}
where $d\Omega$ is the differential solid angle of the Sun from the Earth perspective, $k_B$ is the Boltzmann constant, and $c$ is the speed of light.

In line with previous reports \citep{Linsky1973,Krotikov1987,Hafez2014,Kallunki2018}, $T_{NM}(\nu)$ is set to increase linearly from $\sim$243.97 K to 248.43 K when the frequency increases from 35 to 40 GHz. Five new Moon observations were carried out by CBS dated on 2020 September 18, 2020 October 17, 2021 August 9, 2021 September 7, and 2022 May 2. According to these data, the ratio $\beta$ is always in the range of 43 to 50, being consistent with that derived by \citet{Kuseski1976}. Note that according to the long-term monitoring observations of NoRP\textsuperscript{\ref{fn:1}}, the variation of the quiet Sun flux density hardly exceeds 5\% at 35~GHz during a solar cycle, indicating $\beta$ doesn't change considerably during the above periods. Therefore the average of the new Moon data can be used for calibration. The daily average flux densities of the quiet Sun have been subtracted in our data analysis.

\section{Observations}
\label{section:3}
The following data are used: 1) the CBS radio flux densities at ten frequencies from 35.25 to 39.75~GHz, stepped by 0.5~GHz; 2) the NoRP radio flux densities at 1, 2, 3.75, 9.4, 17 and 35 GHz; 3) the Atmospheric Imaging Assemble \citep[AIA, ][]{Lemen2012} EUV images at 131~\AA{} and 171~\AA{} onboard the Solar Dynamics Observatory \citep[SDO, ][]{Pesnell2012}; 4) the GOES soft X-ray (SXR) data at 0.5 - 4.0~\AA{ and 1 - 8~\AA{}; and 5) the Konus-Wind hard x-ray (HXR) data at 18 - 80 keV, 80 - 328 keV, and 328 - 1300 keV.

\subsection{Event overview}
\label{section3.1}
Figure \ref{fig1} presents the AIA images at 131 and 171~\AA{} during the flare on 2022 April 20 at the southwest limb. The event is partially occulted by the solar disk, and starts at 03:41~UT. At $\sim$03:45~UT, two sets of arcade loops emerge and ascend rapidly according to the 131~\AA{} data (marked by black arrows in Figure \ref{fig1}(a) and (c)). They are not observed at 171~\AA{}, indicating they are high-temperature structures. Around 03:53~UT, the structures expand and brighten. After $\sim$03:54~UT, the AIA data get saturated, and around 03:57~UT the flare reaches its peak. At $\sim$04:00~UT, the large-scale loop system erupts (see the red arrow in Figure \ref{fig1}(f)).

The GOES SXR fluxes (solid lines in Figure \ref{fig2}(a)) and their corresponding time derivatives (dashed lines) reach their maxima at 03:57:25~UT and 03:55:04~UT, respectively, with the latter being earlier by about 2 min. Before their peak, the temporal profiles manifest a first-gradual-then-rapid increase during the impulsive phase, with the turning points at $\sim$03:54:24~UT. The microwave flux densities observed by NoRP (Figure \ref{fig2}(b)) and CBS (Figure \ref{fig2}(c)) show profiles similar to the SXR time derivatives.

In Figure \ref{fig3}, we show other data sets around the flare peak (03:52~UT - 03:58~UT). From top to bottom, the panels show (a) the CBS dynamic spectrum, the CBS (b) and the NoRP (c) microwave flux densities, (d) the GOES SXR time derivatives and the Konus-Wind HXR flux, (e) the microwave polarization degree and the ratio of the NoRP to CBS data around 35~GHz. The dynamic spectrum presents continuum emission from 35 to 40~GHz (see Figure \ref{fig3}(a)). Two enhancements appear around 03:53 and 03:55~UT. The event can be split into the pre-impulsive (03:52:00 - 03:54:24~UT), impulsive (03:54:24 - 03:55:29~UT), and decay (03:55:29 - 03:58:00~UT) stages.

During the pre-impulsive stage, the flux density from 35 to 40 GHz reaches $\sim$50~SFU, along with the eruption of the hot arcade (Figure \ref{fig1}(a) and (c)), while no enhancement of HXR appears. The amplitude of the data fluctuation is $\sim$20 - 30~SFU, and this can be taken as the effective error of measurements of CBS. It is much smaller than the flux density during the impulsive stage. During this stage, the microwave flux densities above 35~GHz reach up to thousands of SFU (see Figure \ref{fig3}(b) and Figure \ref{fig3}(c)). Three distinct local peaks can be identified. The HXR flux of 80 - 328~keV (orange line in Figure \ref{fig3}(d)) shows a similar profile to the microwave data, with the same local peaks. The low-energy lightcurve of HXR (18 - 80~keV, red line in Figure \ref{fig3}(d)) is similar to those of the SXR, while the high-energy lightcurve (328 - 1300~keV, green line in Figure \ref{fig3}(d)) manifests only one significant peak around the flare peak. In the decay stage, all emissions decline rapidly in intensity, and the microwave and high-energy HXR ($>$80 keV) intensities decrease faster than other sets of data.

As seen from Figure \ref{fig3}(c), during the impulsive stage the flux densities at 35~GHz (red) are larger than those at 17~GHz (blue), suggesting that the turnover frequency is above 17~GHz. The flux density ratio of the NoRP data at 35~GHz to the CBS data at 35.25~GHZ (black line in Figure \ref{fig3}(e)) lies between $\sim$0.7 and 1.15 for flux densities above 100~SFU. The relative difference of data magnitudes of CBS and NoRP is $\le15\%$ during the impulsive stage.

\subsection{Microwave spectral fitting}
\label{section3.2}
The nonthermal microwave spectra can be fitted with the following function \citep[see, e.g.][]{Asai2013}
\begin{equation}
I(\nu)=I_{t0}(\frac{\nu}{\nu_{t0}})^{\alpha_{tk}}(1-\rm{exp}(-(\frac{\nu}{\nu_{t0}})^{\alpha_{tn}-\alpha_{tk}})),
\label{eq:fitting}
\end{equation}
where $\nu_{t0}$, $I_{t0}$, $\alpha_{tk}$, $\alpha_{tn}$, and $I(\nu)$ are the fitting parameters representative of the turnover frequency, the flux density at the turnover frequency, the optically-thick and -thin spectral indices, and the flux density at $\nu$, respectively. The ``real" value of $\nu_{t}$ ($I_{t}$) given by the fitted spectra is slightly different from the above fitting parameter $\nu_{t0}$ ($I_{t0}$).

To eliminate the systematic error between the two instruments, we cross-calibrate the CBS data by multiplying their values with the flux density ratio as shown in Figure \ref{fig3}(e). Figure \ref{fig4} presents the fitted spectra and the corresponding parameters for 13 selected moments ($T_1 - T_{13}$). During the pre-impulsive stage, the fitted spectra manifest a typical gyrosynchrotron spectrum (see the green line in Figure \ref{fig4}(a)), with a turnover frequency of 11.1~GHz. The turnover frequency $\nu_{t}$ goes above $\sim$15~GHz, and the corresponding intensity $I_{t}$ goes above 1000~SFU during most of the impulsive stage extending from $T_2$ to $T_{10}$ (Figure \ref{fig4}(a) - (c)). The maximum flux density is $\sim$9300~SFU, obtained with the middle and major peak around $T_6$. During the decay stage (Figure \ref{fig4}(d)), both $\nu_{t}$ and $I_{t}$ decrease with the declining flux density, for instance, the paired values of $\nu_{t}$ (GHz) and $I_{t}$ (SFU) equal to (25.2, 709) at 03:55:36~UT ($T_{11}$), (23.9, 464) at 03:55:46~UT ($T_{12}$), and (23.0, 282) at 03:55:56~UT ($T_{13}$).

Figures \ref{fig5}(a) and \ref{fig5}(b) present the evolution of spectral parameters ($\alpha_{tk}$, $\alpha_{tn}$, $\nu_t$, and $I_t$). The spectral index $\alpha\rm{_C}$ according to the CBS data within 35 - 40 GHz has been overplotted in Figures \ref{fig5}(a). Large uncertainties exist for spectral fitting with $\nu_t$ $\geq$35~GHz, and therefore at such moments the values of $\alpha_{tn}$, $\nu_t$, and $I_t$ given by the fittings are not shown in the figure. No evident variation of the optically-thick index $\alpha_{tk}$ is observed from 03:54 UT to 03:56 UT, while the optically-thin index $\alpha_{tn}$ changes significantly. Note that $\alpha_{tn}$ changes similar to $\alpha\rm{_C}$. Both data reveal that $\nu_t$ is above 20 GHz for most of the first (yellow shadow) and third (orange shadow) peaks of the impulsive stage. At the start of the middle peak, the CBS index $\alpha\rm{_C}$ increases sharply from negative ($\sim-2$) to positive ($\sim1$), and remains to be positive for $\sim$15s. This means that the turnover frequency $\nu_t$ is larger than $\sim$ 35-40 GHz during this period.

During the decay stage of this peak, the CBS index $\nu\rm{_C}$ returns to be negative ($\sim$-2 to -1). As seen from Figure \ref{fig5}, the CBS spectra present a rapid-softening-then-rapid-hardening profile in between two neighbouring peaks, likely due to the rapid and intermittent energy release associated with the peaks.

In the decay stage, gradual hardening is observed from the temporal profiles of both $\alpha\rm{_C}$ and $\alpha_{tn}$. The difference between two sets of spectral indices is close to 1. This difference is due to the fact that $\alpha_{tn}$ represents the fitted slope at the high-frequency limit, while $\nu\rm{_C}$ is given by the CBS measurement within 35 - 40~GHz.

Figure \ref{fig5}(b) presents a nice correlation between $I_t$ and $\nu_t$ during the whole impulsive stage. The correlation can also be seen from Figure \ref{fig5}(c), which shows that $I_t$ ($\nu_t$) can be well fitted with a power-law relation ($I_t \propto \nu_t^{4.8}$). The correlation coefficient is about 0.82.

\section{SUMMARY AND DISCUSSION}
\label{sec:4}
We reported the first flare observation of the newly-built Chashan Broadband Solar millimeter spectrometer (CBS) that has begun its routine observation from 35 to 40~GHz since 2020. The CBS data are first calibrated with the new Moon observations, and then cross-calibrated with the simultaneously \textbf{NoRP} data at 35~GHz. The flare is of special interest due to its strong millimeter emission and the high turnover frequency ($>$20 GHz) of the spectra during the impulsive stage. Such events have not been well-studied due to the large data gap beyond 17 GHz.

Three distinct local intensity peaks exist during the impulsive stage. The middle peak is the strongest one, with the largest flux density reaching $\sim$9300~SFU at 35 - 40~GHz. The gyrosynchrotron turnover frequency ($\nu_t$) is above 35 - 40 GHz for this major peak, according to the positive spectral indices of the CBS data there. The turnover frequency $\nu_t$ is larger than 20~GHz for most of the other two peaks. A systematic rapid-hardening-then-softening trend can be identified from the CBS data for each peak, agreeing with the spectral fittings with the combined NoRP-CBS data for moments with the turnover frequency $\nu_t<$35~GHz. We found $I_t$ correlates well with $\nu_t$ according to the power-law relation $I_t \propto \nu_t^{4.8}$ during the impulsive stage with $\nu_t<$35~GHz. During the decay stage, both the CBS spectral index $\nu\rm{_C}$ and the fitted optically-thin spectral index $\alpha_{tn}$ present a gradual hardening trend.

The nice power-law correlation between $I_t$ and $\nu_t$ has been reported earlier in a statistical study with 38 flares observed by the OVSA with the turnover frequencies ranging from 3 to 16 GHz \citep{Melnikov2008}. They studied the variation of the event spectra with $\nu_t$ during both the rising and decay stages. They revealed a similar power-law correlation of $I_t$ versus $\nu_t$ with the power-law indices being $\sim$3.7 during the rising and $\sim$1.75 during the decay stage on average. These indices are smaller than those obtained here ($\sim 4.8$). Their result was obtained for events with turnover frequencies much lower than that reported here.

The HXR data of Konus-Wind are consistent with the spectral evolution during the impulsive stage. The lightcurves (80-328 keV and 328-1300 keV, orange and green lines in Figure \ref{fig3}(d)) peak around 03:54:46 UT, earlier by $\sim$ 30s than the peak of the lightcurve of 18-80 keV. This indicates that the rapid precipitation of energetic electrons towards the lower corona leads to the rapid softening of the microwave and HXR spectra \citep{Melrose1976}. The overall rapid-hardening-then-softening trend of the CBS spectra for each peak can also be understood by the rapid and intermittent injection-then-loss process. The observed gradual hardening of the microwave spectra during the decay stage is consistent with some earlier reports \citep{Melnikov1998,Ning2007,Asai2013}.

Earlier studies were limited to events with a much smaller turnover frequency either being $<$9.4 GHz \citep{Ning2007} or $<$17 GHz \citep{Asai2013}, to ensure two data points are available so the optically-thin spectra can be inferred. Here the spectral turnover frequency and other fitting parameters can be better constrained with the CBS data covering the range of 35 - 40 GHz. This is true for moments when the turnover frequency is lower than 35 GHz. During the second peak of the impulsive stage, valuable information can still be inferred according to the unique CBS data though the exact turnover frequency cannot be determined. Such data are available for the the first time in the millimeter observations of solar radio bursts. Data with a broader spectral coverage are still demanded to better understand flares with a high-turnover frequency.


\acknowledgments
This work is supported by the grants of National Natural Science Foundation of China (11790303 and 42127804).



\begin{figure}
\centering
\epsscale{.9}
\plotone{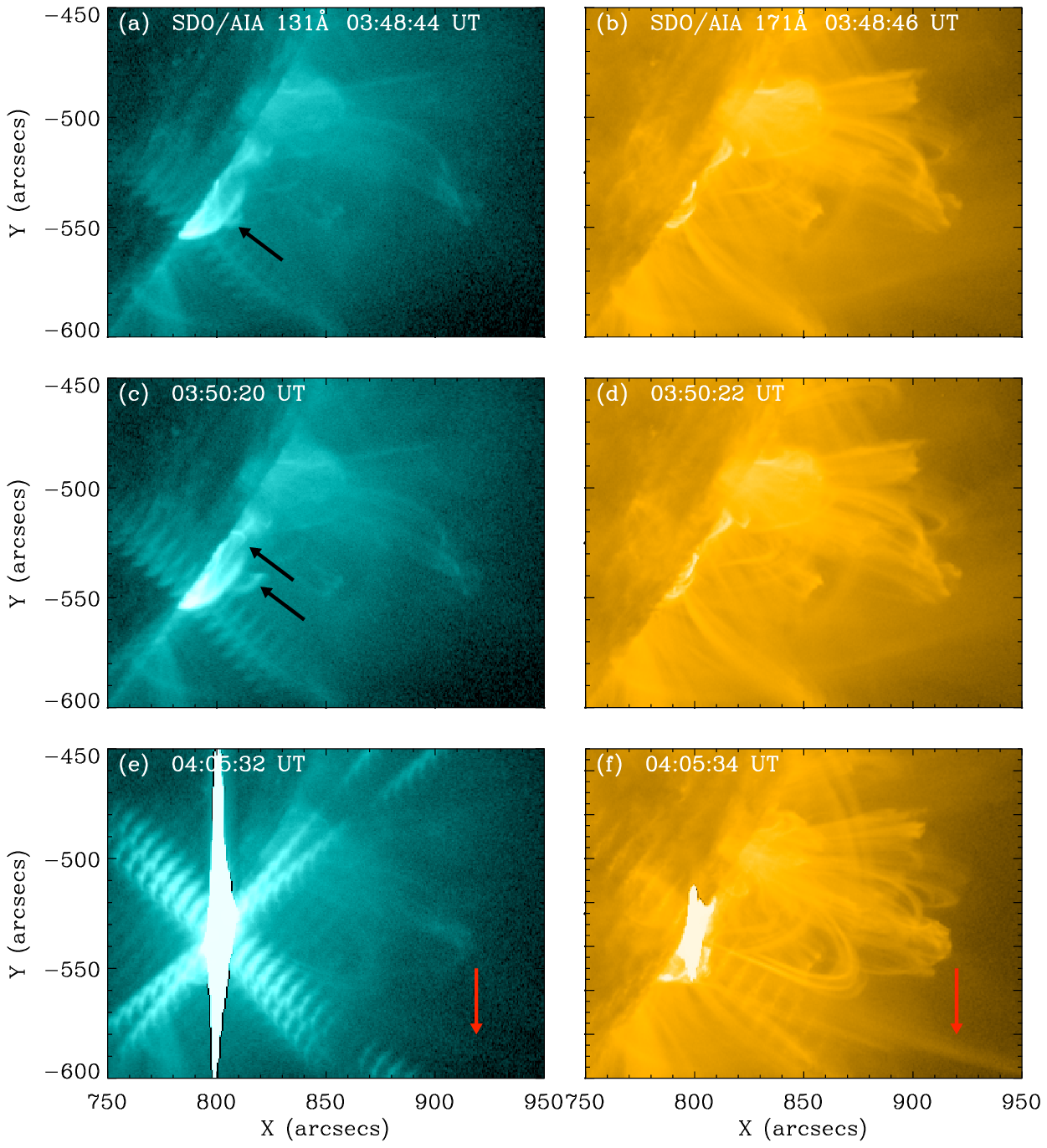}
\caption{The AIA 131~\AA{} (left) and 171~\AA{} (right) images of the partially-occulted X2.2 flare. The black arrows in (a) and (c) point at the expanding arcade loops during the pre-impulsive stage, and the red ones in (e) and (f) point at the erupting large-scale loop system.
}
\label{fig1}
\end{figure}

\begin{figure}
\centering
\epsscale{.9}
\plotone{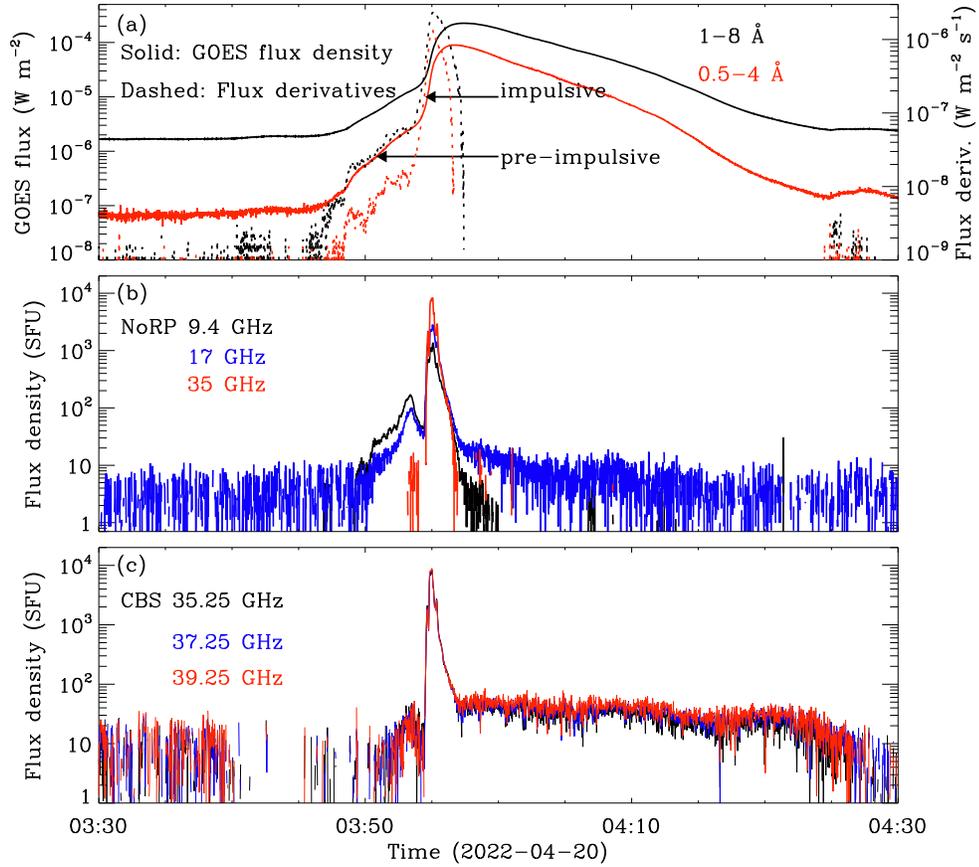}
\caption{Overview of the X2.2 flare observed by GOES, NoRP, and CBS on April 20 2022: (a) the SXR fluxes (solid) and their time derivatives (dashed) observed at 1-8~\AA{} (black) and 0.5-4~\AA{} (red) by GOES; (b) the microwave flux density observed by NoRP at 9.4 (black), 17 (blue), and 35 (red)~GHz; (c) the microwave flux density observed by CBS at 35.25 (black), 37.25 (blue), and 39.25 (red)~GHz.}
 \label{fig2}
\end{figure}
\begin{figure}
\centering
\epsscale{1.9}
\plotone{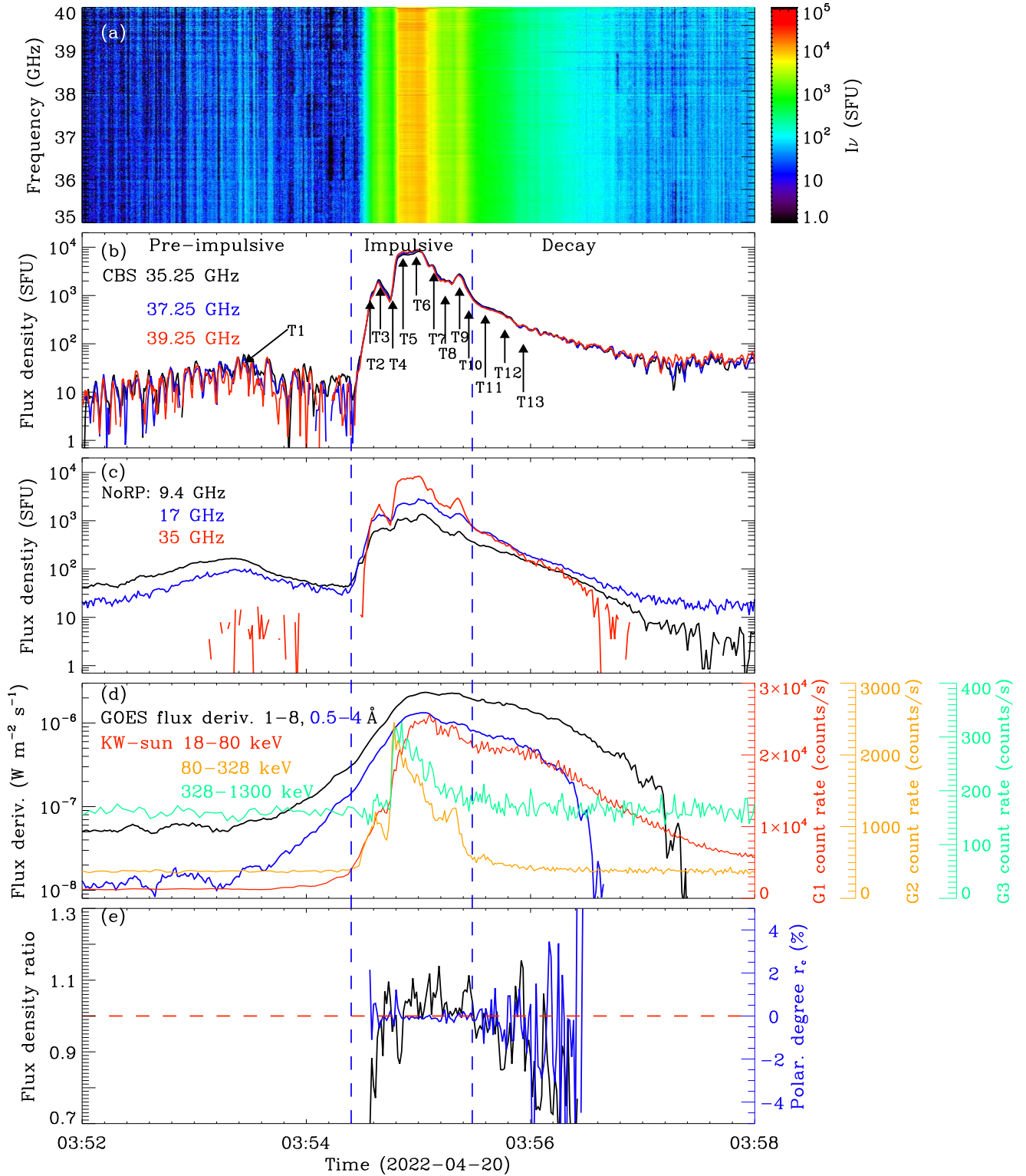}
\caption{Various data sets obtained from 03:52~UT to 03:58~UT: (a) the CBS dynamic spectrum; (b) the CBS microwave flux desity; (c) the NoRP microwave flux density, (d) the GOES SXR derivatives and Konus-Wind (KW-Sun) HXR profiles, and (e) the polarization degree (blue) of the NoRP data at 35 GHz, and the ratio (black) of the NoRP data at 35~GHz to the CBS data at 35.25~GHz. Vertical lines mark the separations between the pre-impulsive, impulsive, and decay stages. Vertical arrows in panel (b) denote 13 representative moments of the flare evolution, spectra at these moments are presented in Figure \ref{fig4}.}
\label{fig3}
\end{figure}
%
%

\begin{figure}
\centering
\epsscale{1.0}
\plotone{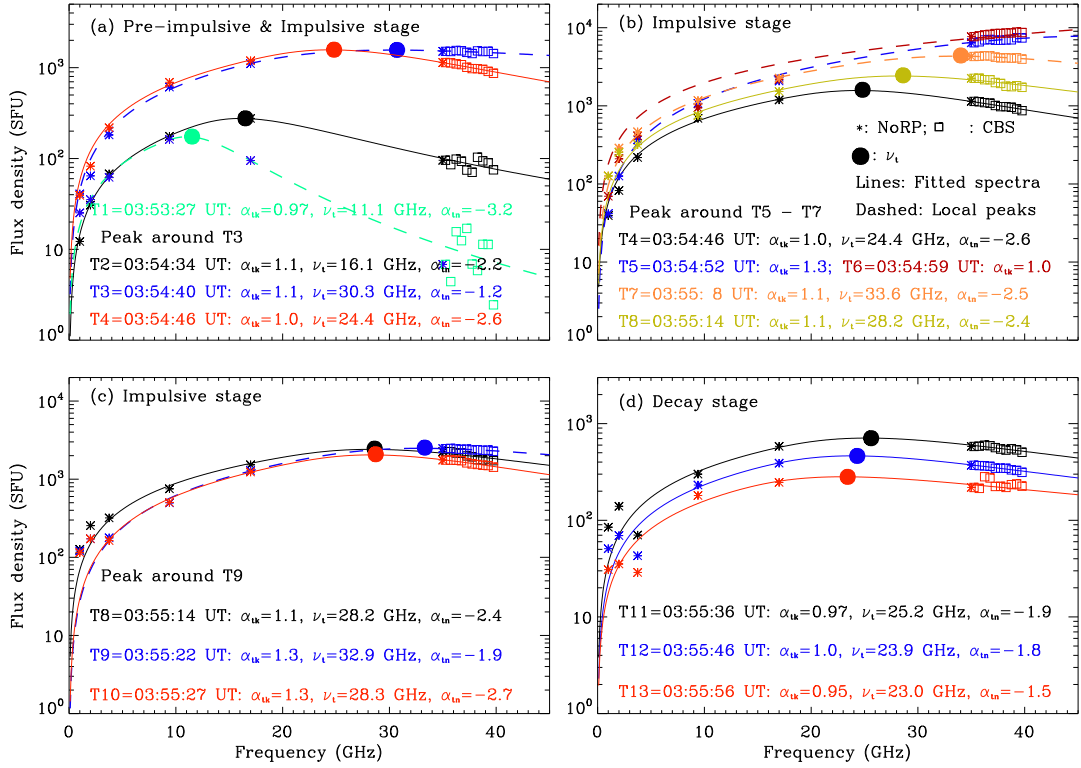}
\caption{Spectra obtained by fitting the combined NoRP and CBS data at selected moments of T$_1$ - T$_{13}$. The asterisks and squares represent the NoRP and CBS data, repectively. The spectral parameters, including the spectral indices $\alpha_{tk}$, $\alpha_{tn}$, and $\nu_{t}$ are given in each panel. The solid circles denote the location of $\nu_t$.
}
\label{fig4}
\end{figure}
\begin{figure}
\centering
\epsscale{.75}
\plotone{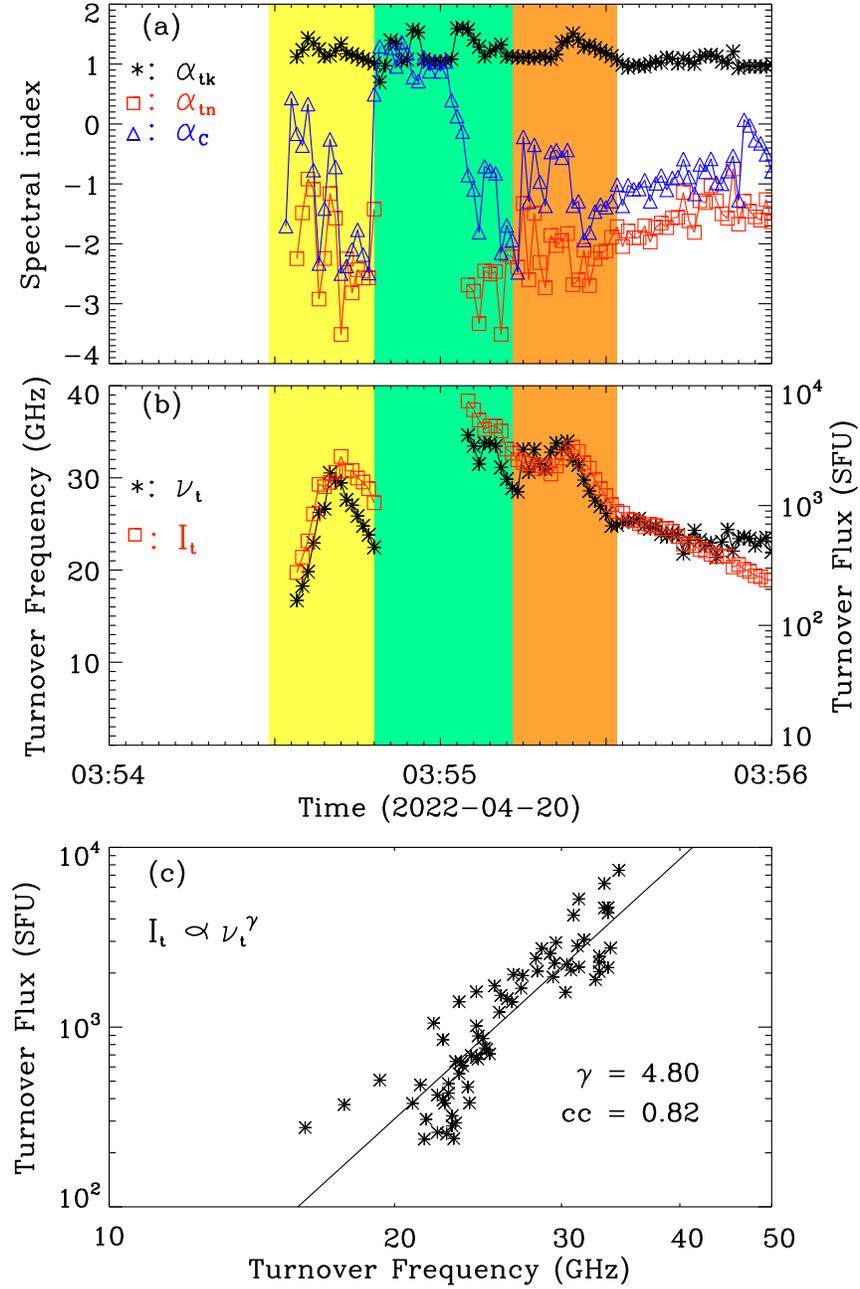}
\caption{Evolution of the spectral parameters from 03:54 UT to 03:56~UT: (a) the fitted optically-thick and -thin spectral indices ($\alpha_{tk}$ and $\alpha_{tn}$) are plotted as black asterisks and red squares, repectively. The CBS spectral index ($\alpha\rm{_C}$) is overplotted as blue triangles; (b) the turnoever frequency and the corresponding intensity ($\nu_{t}$ and $I_{t}$) are plotted as black asterisks and red squares, repectively; (c) the power-law fitting of $I_t$ versus $\nu_t$ with the combined NoRP and CBS data at moments with the turnover frequency being less than 35~GHz.
}
 \label{fig5}
\end{figure}

\end{document}